\documentclass[reprint,amsmath,amssymb,aps,prl,superscriptaddress,oneside]{revtex4-2}

\usepackage{graphicx}
\usepackage{color}
\usepackage{epstopdf}
\usepackage[usenames,dvipsnames]{xcolor}
\usepackage{bm}
\usepackage{paralist}
\usepackage{amsmath}
\usepackage{amsthm}
\usepackage{amssymb}
\usepackage{pstool}
\usepackage[percent]{overpic}
\usepackage{rotating}
\usepackage{lipsum}
\usepackage[normalem]{ulem}
\usepackage{titlesec}
\usepackage{fouriernc}
\usepackage[T1]{fontenc}
\usepackage{enumitem}
\usepackage{cancel}
\usepackage[small]{caption}
\usepackage[skip=1pt]{subcaption}

\setlist[itemize]{noitemsep, nolistsep}

\definecolor{Xred}{HTML}{B31B1B}

        \skip\footins 15pt
\usepackage{scalerel}
\usepackage{tikz}
\usetikzlibrary{svg.path}

\definecolor{verylightgray}{HTML}{F3F3F3}
\definecolor{notsoverylightgray}{HTML}{E3E3E3}

\definecolor{orcidlogocol}{HTML}{A6CE39}
\tikzset{
  orcidlogo/.pic={
    \fill[orcidlogocol] svg{M256,128c0,70.7-57.3,128-128,128C57.3,256,0,198.7,0,128C0,57.3,57.3,0,128,0C198.7,0,256,57.3,256,128z};
    \fill[white] svg{M86.3,186.2H70.9V79.1h15.4v48.4V186.2z}
                 svg{M108.9,79.1h41.6c39.6,0,57,28.3,57,53.6c0,27.5-21.5,53.6-56.8,53.6h-41.8V79.1z M124.3,172.4h24.5c34.9,0,42.9-26.5,42.9-39.7c0-21.5-13.7-39.7-43.7-39.7h-23.7V172.4z}
                 svg{M88.7,56.8c0,5.5-4.5,10.1-10.1,10.1c-5.6,0-10.1-4.6-10.1-10.1c0-5.6,4.5-10.1,10.1-10.1C84.2,46.7,88.7,51.3,88.7,56.8z};
  }
}

\newcommand\orcidicon[1]{\href{https://orcid.org/#1}{\mbox{\scalerel*{
\begin{tikzpicture}[yscale=-1,transform shape]
\pic{orcidlogo};
\end{tikzpicture}
}{|}}}}

\newtheorem*{theorem*}{Theorem}
\newtheorem*{corollary*}{Corollary}
\newtheorem*{lemma*}{Lemma}

\newtheorem*{assumption*}{Assumption}

\newtheorem*{principle*}{Principle}

\usepackage[colorlinks=true,linkcolor=Magenta,citecolor=Magenta,breaklinks=true]{hyperref}
\def\calE{\mathcal{E}}
\def\calT{\mathcal{T}}
\def\calF{\mathcal{F}}
\def\calG{\mathcal{G}}
\def\supp{\text{supp}}

\begin{document}
\title{Spectroscopic classification of non-ergodic populations}

\author{Nicolas G. Underwood\,\orcidicon{0000-0003-4803-2629}}
\email{NUnderwood@lincoln.ac.uk}
\author{Fabien Paillusson\,\orcidicon{0000-0002-5740-3463}}
\email{FPaillusson@lincoln.ac.uk}
\affiliation{School of Engineering and Physical Sciences, University of Lincoln, Brayford pool, Lincoln, LN6 7TS, United Kingdom}
\date{\today}

\makeatletter
\renewcommand\frontmatter@abstractwidth{5.6in}
\makeatother

\begin{abstract}
\hspace{-\parindent}%
Non-ergodicity impacts statistical inference in a diverse range of disciplines inside and outside of physics.
However the concept of ergodicity is used inconsistently, and may refer to several nonequivalent notions.  
To help address this, we first identify and clarify the relationship between three major interpretations of ergodicity.
We then introduce a method of spectral analysis of non-ergodicity which may be performed using data alone, and so may be applied in both numerical and empirical contexts.
This may be used to identify, quantify, and classify non-ergodic populations within an ergodic decomposition.
This is demonstrated with an application to the Kob-Andersen kinetically constrained lattice glass model.
\end{abstract}

\maketitle

\titleformat{\section}[runin]{\normalfont\itshape\bfseries}{}{0em}{}[.---]
\titlespacing{\section}{\parindent}{0pt}{0pt}

\setlength{\belowdisplayskip}{3pt} \setlength{\belowdisplayshortskip}{3pt}
\setlength{\abovedisplayskip}{3pt} \setlength{\abovedisplayshortskip}{3pt} 

\section{Introduction}
The extent to which inferences based on ensemble statistics may be applied to a single system over time (and vice versa) impacts a broad range of scientific topics outside of physics,
with notable discussions currently taking place, for instance, on the temporal variability of climate \cite{climate9,climate10,climate8,climate1,climate2,climate11};
the interpretation of clinical data in medicine \cite{medical1,medical2,medical3,medical4,medical5}
and psychology \cite{psychology3,psychology1,psychology4,psychology2,psychology7}; 
and the validity of economic models \cite{economics1,economics2,economics3,economics4,economics6}.
In physics, this statistical equivalence is traditionally established as a consequence of the dynamical property of \emph{ergodicity},  
which received its first mathematically strict treatment in the form of metric indecomposability/transitivity by Birkhoff and von Neumann \cite{B31,vN32}, and has since been extended and generalized by many others \cite{kallenberg,C19,Gallager,K11}.
However this notion (often simply called ergodicity at the suggestion of von Neumann \cite{ergodic_history}) has been difficult to prove even for relatively simple physical systems (e.g.~dynamical billiards \cite{hard_spheres_history,D14}, the FPUT model \cite{BCP13,BP23}).
In disciplines where the dynamics may be many dimensional, complex, or unknown and empirically investigated, any such proof is often out of reach.
Perhaps for this reason, ergodicity has branched into various related but nonequivalent notions. 
We broadly divide these into three categories we refer to as:
\emph{metric indecomposability} (MI), 
\emph{the equality of averages} (EoA), 
and \emph{the equality of distributions} (EoD).
The choice of notion used is often discipline dependent; Molenaar's \cite{psychology3} EoD-style notion of ergodicity has become an accepted standard in medicine and psychology ~\cite{medical1,medical2,medical3,medical4,medical5,psychology1,psychology4,psychology2,psychology7},
while a growing number of economics studies have adopted Peters' \cite{economics1} EoA notion \cite{economics2,economics3,economics4,economics6}.


\section{Metric indecomposability (MI)}
Mathematics textbooks \cite{Walters,VO16} usually describe MI as a property of a measure preserving dynamical system
$(\Omega,\mathcal{B},m,U)$, represented by a state space $\Omega$;
a $\sigma$-algebra $\mathcal{B}$ over $\Omega$;
a probability measure $m:\mathcal{B}\to [0,1]$;
and a dynamical transformation $U:\Omega\to \Omega$ that is measure preserving, $m(U^{-1}\omega)=m(\omega)$ for all $\omega\in \mathcal{B}$.
As this setting can be a barrier to those unfamiliar with measure theory, we restrict our discussion of such systems to a continuous state/phase space $\Omega$ with coordinates $x$, a continuous time dynamics $x\to U_tx$, and an ensemble density $\mu(x)$ which is finite for all $x$, and that is related to the measure by $m(\omega)=\int_\omega\mu(x)\mathrm{d}x$.
This density is stationary under dynamical evolution by virtue of the presumptive measure preserving property. 
We furthermore assume all subspaces $\omega\subseteq\Omega$ we refer to are members of $\sigma$-algebra $\mathcal{B}$.

In this deterministic context, system trajectories are confined to so-called \emph{invariant volumes}, meaning subspaces $\omega\subseteq\Omega$ that are static under dynamical evolution, $U_t^{-1}\omega=\omega$.
These are said to be \emph{decomposable} if they may be divided into smaller (measure nonzero) invariant volumes. 
MI is the condition that $\supp(\mu)$ (the support of $\mu(x)$, an invariant volume with unit measure) is not decomposable. 
Equivalently, any trajectory with initial state $x_0$ chosen randomly within $\supp(\mu)$ will in the limit $t\to\infty$ spend a proportion of its time in any volume $\omega$ equal to $m(\omega)=\int_\omega\mu(x)\mathrm{d}x$ \cite{VO16}, and so sample the space as envisioned by Boltzmann/Maxwell \cite{G16,Ehrenfests}.
(A common misconception appears to be that MI ensures dynamical relaxation of a nonequilibrium density to $\mu(x)$, which instead follows from the stronger property of \emph{mixing} \cite{BLP2014}.)

Analogous MI conditions apply in stochastic contexts \cite{Gallager,G09,K11,H08}. 
For instance, below we illustrate our spectral analysis method with a discrete-time Markov chain $\text{Prob}(j\to i)=T_{ij}$ on a countably large \footnote{For instance, in Fig.~\ref{configs_figure} there are $N=C^{100}_{81}\sim10^{20}$ states.} but finite space, $i,j\in\Omega=\{1,..,N\}$, wherein the role of indecomposable invariant volumes is played by \emph{recurrent classes} of states.
That is, subsets of the space within which a system in any state possesses a non-vanishing probability of reaching any other state after some finite number of time steps. 
In this context the MI condition takes the form of \emph{irreducibility}, meaning there exists only a single recurrent class of states.
(There may also exist transient states \cite{Gallager}.)
In contrast to measure preserving dynamical systems, where a stationary ensemble density $\mu(x)$ is axiomatized \footnote{Under certain conditions the existence of an invariant measure may be argued for deterministic systems with the Krylov-Bogolyubov theorem.}, for Markov chains (in which ensemble densities $p_i$ evolve as $p_i\to \sum_j T_{ij}p_j$) the existence of at least one ensemble density $\mu_i$ that is stationary, $\mu_i=\sum_jT_{ij}\mu_j$, follows from either the Perron-Frobenius theorem or Brouwer's fixed-point theorem applied to the convex set of densities.


\section{Equality of Averages (EoA)}
Context notwithstanding, the ostensible role of the various ergodic theorems \cite{kallenberg} is to establish the sufficiency of MI to ensure the equality of ensemble and time averages of a state function $f$.
For the above measure preserving dynamical system, this equality may be expressed
\begin{align}\label{EoA}
\lim_{t_f\to\infty}\frac{1}{t_f}\int_0^{t_f}f(U_tx_0)\,\mathrm{d}t=
\int_\Omega f(x)\mu(x)\,\mathrm{d}x,
\end{align}
and applies \emph{for all} (in this case integrable \cite{ergodic_history}) state functions $f(x)$, and almost all initial conditions $x_0$.
By \emph{almost all} $x_0$ it is meant that if $x_0$ is chosen at random from within the indecomposable invariant volume $\supp(\mu)$, then EoA Eq.~\eqref{EoA} holds with unit probability.

For an MI Markov chain, after $t\in\mathbb{N}$ time steps a system initially in state $j$ is distributed as $(T^t)_{ij}$, however this only converges to $\mu_i$ in the limit $t\to\infty$ if the recurrent class of states is \emph{aperiodic} \cite{Gallager} (similar to mixing for deterministic systems).
Nevertheless, the time average of a state function $f_i$ over any trajectory is a Ces\`{a}ro average whose almost sure convergence follows from the strong law of large numbers, so that in analogy to Eq.~\eqref{EoA}, MI establishes the EoA
\begin{align}\label{EoA_Markov}
\lim_{t_f\to\infty} \frac{1}{t_f+1}\sum_{t=0}^{t_f}\sum_i f_i(T^t)_{ij}
=\sum_i f_i\mu_i
\end{align}
for all states $j\in \Omega$.
For brevity, we do not state the corresponding Markov chains mathematics going forwards. 


\section{Equality of distributions (EoD)}
EoD follows from the use of an indicator function $\mathbf{1}_\omega(x)$, which is equal to unity for $x\in\omega$ and vanishes otherwise.
By setting $f(x)\to \mathbf{1}_\omega(x)$ in Eq.~\eqref{EoA} with $\omega$ equal to the region of the state space $\Omega$ in which the value of a dummy variable $y$ is less than original function $f(x)$, i.e.~$\omega=\{x|f(x)\leq y\}$, one may pick out the portion of the trajectory and ensemble in which the state function is below this value so that
\begin{align}\label{EoD}
\calT[f](y,x_0):=&\lim_{t_f\to\infty}\frac{1}{t_f}\int_0^{t_f}\mathbf{1}_{\{x|f(x)\leq y\}}\left(U_tx_0\right)\mathrm{d}t\nonumber\\
=\calE[f](y):=&\int_{\left\{x|f(x)\leq y\right\}}\mu(x)\mathrm{d}x,
\end{align}
defining the equality of trajectory and ensemble (cumulative) distributions, $\calT[f](y,x_0)$ and $\calE[f](y)$. 


\section{MI $\Longrightarrow$ EoD $\Longrightarrow$ EoA}
While ergodic theorems show the sufficiency of MI to establish EoA for all functions, it is also possible to prove the necessity and so establish an equivalence. See for instance Proposition 4.1.3 of Ref.~\cite{VO16}.
Thus,
\begin{align}
\text{MI }\Longleftrightarrow\text{ EoA for all } f \Longrightarrow \text{ EoD for all }f.\nonumber
\end{align}
However in practical circumstances, verification that EoA holds \emph{for all functions} is clearly out reach, and potentially more difficult than proving MI.
So instead we take EoA and EoD to be properties assigned on a function-by-function basis as for instance is the convention of Ref.~\cite{economics1}. 
Then, since for any particular function EoD is a stronger condition than EoA, the three conditions form the implicative hierarchy
\begin{align}
\text{MI }\Longrightarrow\text{ EoD for a given } f \Longrightarrow \text{ EoA for the same }f.\nonumber
\end{align}

As by contraposition $\neg$EoA $\Longrightarrow$ $\neg$EoD $\Longrightarrow$ $\neg$MI, it is possible to formulate significance tests to rule out MI by holding either EoA or EoD as a null hypothesis.
In each case, tests may be performed with the same data set (a sample of i.i.d.~function values both over a trajectory and over an ensemble).
However as EoD is a stronger condition, it is not unreasonable to expect EoD based tests to be typically more powerful, and it is simple to conceive of circumstances in which an EoD based test provides a significant result, while an EoA based test does not. 
Additionally, a number of established tests of distributional equality are non-parametric \cite{F49,P76,A62}, meaning they may be performed without any assumptions made on the underlying distributions $\calT[f](y,x_0)$ and $\calE[f](y)$.
See for instance Ref.~\cite{PF12} in which EoD and thus MI was confidently ruled out for some amplitudes of a tapped granular system.


\section{Non-ergodicity and ergodic decomposition}
\begin{figure}[h]
\vspace{-05pt}
\includegraphics[width=\columnwidth]{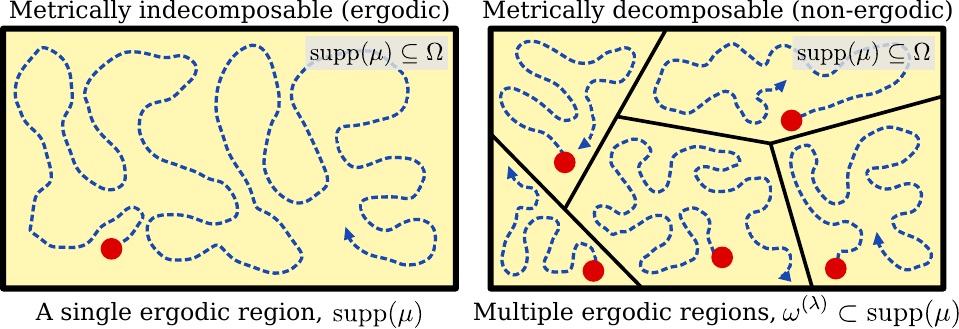}
\caption{In a non-MI (non-ergodic) system, the state space may be decomposed into individually ergodic regions. By Eq.~\eqref{decomposed_spectrum}, each ergodic region represented in an ensemble contributes a spectral line to an ergodic spectrum.}
\vspace{-15pt}
\end{figure}

\begin{figure*}[ht]
\setlength{\fboxsep}{0pt}%
\setlength{\fboxrule}{1pt}%
\includegraphics[width=\textwidth]{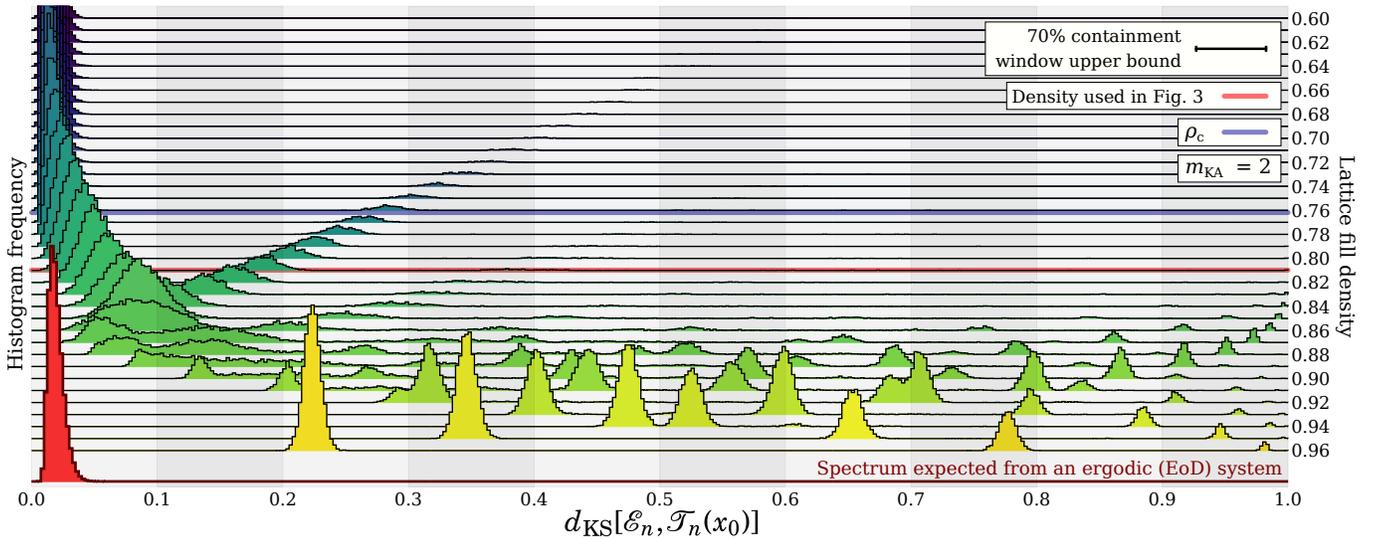}
\vspace{-15pt}
\caption{Ergodic spectra for different densities of the 2d KA model on a 10$\times$10 lattice with sample size $n=4000$ (controlling spectral resolution) and super sample size $n_\text{sup}=10,000$ (controlling noise). 
Lines represent ergodic regions/populations via Eq.~\eqref{decomposed_spectrum}.
}\label{spectra}
\vspace{-20pt}
\end{figure*}

Systems that are MI wrt density $\mu(x)$ are commonly referred to as being \emph{$\mu$-ergodic}.
When this is not the case, the invariant volume $\supp(\mu)$ may be \emph{decomposed} into disjoint invariant volumes, the smallest of which are indecomposable \footnote{See Chp. 5 of Ref.~\cite{VO16} for a rigorous discussion.}.
Denoting these $\omega^{(\lambda)}$, formally we write $\supp(\mu)=\bigcup_\lambda\omega^{(\lambda)}$ and $\omega^{(\lambda)}\cap\omega^{(\xi)}=\emptyset$ for $\lambda\neq \xi$, though for a continuous $\Omega$ label $\lambda$ may take uncountable values (see e.g.~Ref.~\cite{VO16}). 
Given a stationary ensemble density $\mu^{(\lambda)}(x)$ defined upon an indecomposable volume $\omega^{(\lambda)}$, so that $\supp(\mu^{(\lambda)})=\omega^{(\lambda)}$, the dynamics is then by construction MI on this subspace, or $\mu^{(\lambda)}$-ergodic.
Furthermore, each $\omega^{(\lambda)}$ admits only a single unique density $\mu^{(\lambda)}(x)$.
To see why, suppose the dynamics did admit two distinct stationary densities $\mu^{(\lambda)}(x)$ and $\nu^{(\lambda)}(x)$ with support $\omega^{(\lambda)}$.
As by construction the dynamics is MI wrt both, by Eq.~\eqref{EoA} a trajectory average with a randomly chosen initial condition $x_0\in\omega^{(\lambda)}$ equates to both $\int_\Omega f(x)\mu^{(\lambda)}(x)\mathrm{d}x$ and $\int_\Omega f(x)\nu^{(\lambda)}(x)\mathrm{d}x$.
Taking $f(x)=\mathbf{1}_\omega(x)$ results in $\int_\omega \mu^{(\lambda)}(x)\mathrm{d}x=\int_\omega \nu^{(\lambda)}(x)\mathrm{d}x$, which holds for arbitrary $\omega$, and so which we take to mean that $\mu^{(\lambda)}(x)=\nu^{(\lambda)}(x)$, thus reaching the intended contradiction.
As a corollary, any stationary density $\mu(x)$ may be uniquely decomposed as  
\begin{align}\label{decomposition}
\mu(x)=\sum_\lambda \alpha^{(\lambda)}\mu^{(\lambda)}(x),
\end{align}
where $\alpha^{(\lambda)}$ is the proportion of ensemble members confined to invariant region $\omega^{(\lambda)}$.
As each $\alpha^{(\lambda)}\geq 0$ and $\sum_\lambda \alpha^{(\lambda)}=1$, the space of stationary densities is formally a convex set with the $\mu^{(\lambda)}$ as its extremal elements.
(Of course, applied to the case of MI/$\mu$-ergodicity, this argument ensures the uniqueness of $\mu(x)$.)


\section{Ergodic and non-ergodic spectra}
To describe non-ergodicity in terms of the departure from EoD \eqref{EoD}, first let $d[\calE,\calT(x_0)]$ be a measure of the statistical distance (for examples see Ref.~\cite{GS02}) between ensemble and trajectory distributions, $\calE[f](y)$ and $\calT[f](y,x_0)$.  
(Going forward we suppress $f$ dependence.) 
Then, given a density of initial states $\mu(x_0)$, the overall probability density of $d[\calE,\calT(x_0)]$ may be expressed 
\begin{align}\label{ideal_profile}
\rho_\text{spec}(z):=\int_\Omega\delta\left[z-d[\calE,\calT(x_0)]\right]\mu(x_0)\mathrm{d}x_0,
\end{align}
where $\delta[...]$ is a Dirac delta, and $z$ is a dummy variable taking values in the range of $d[\calE,\calT(x_0)]$.
We name this density the \emph{ergodicity spectrum} of the system. 

In contrast to a binary test of MI, construction of an ergodicity spectrum may provide information on the ergodic decomposition \eqref{decomposition} of a non-ergodic system.
To see how, note that in a non-ergodic system trajectory distributions $\calT(x_0)$ are invariant on each ergodic region (that is, for all $x_0\in\omega^{(\lambda)}$), so that $\calT(x_0)$ may be replaced with $\calT^{(\lambda)}$ in Eq.~\eqref{ideal_profile}.
Then, using Eq.~\eqref{decomposition}, the spectrum may be re-expressed as a sum over ergodic regions,
\begin{align}\label{decomposed_spectrum}
\rho_\text{spec}(z)=\sum_\lambda\alpha^{(\lambda)}\delta\left[z-d[\calE,\calT^{(\lambda)}]\right].
\end{align}
For an ergodic (MI) system, $d[\calE,\calT(x_0)]=0$ for all $x_0$ (sampled from $\supp(\mu)$), and so a spectrum is a simple delta function or \emph{spectral line} at $z=0$.
For a non-ergodic system with a countable decomposition, $\lambda=1,2,...,n_\text{reg}$, key properties of a spectrum are: 
1) It comprises $n_\text{reg}$ spectral lines corresponding to each ergodic region $\omega^{(\lambda)}$.
2) The location of each line corresponds to the \emph{degree of deviation from EoD} of $\mu^{(\lambda)}$-ergodic trajectories as measured by $d[\calE,\calT^{(\lambda)}]$.
3) The normalization of each line corresponds to $\alpha^{(\lambda)}$, thus indicating the \emph{prevalence} of $\mu^{(\lambda)}$-ergodic trajectories.
Note that $d[\calE,\calT^{(\lambda)}]$ does not necessarily map ergodic regions to distances one-to-one, so lines may overlie one another, however as demonstrated below this multiplicity can be indicative of ergodic regions that are related by a symmetry of the system/state function and so may be considered to belong to a single \emph{ergodic class}. 
Also note that a system with a continuous state space $\Omega$ may possess an uncountable number of ergodic regions, in which case Eq.~\eqref{decomposed_spectrum} describes a continuous rather than a line spectrum.


\section{Noise and spectral resolution in spectrum reconstruction}
In practice, an ergodicity spectrum may be reconstructed by histogramming a \emph{super sample} of $n_\text{sup}$ estimates of the distributional difference $d[\calE,\calT(x_0)]$, with $x_0$ sampled from $\mu(x_0)$.
This introduces two types of error to be managed.
Firstly, the finite size of $n_\text{sup}$ is a source of noise in the histogram, which may be managed by adjusting the histogram bin size in the usual manner.
Secondly, each estimate of $d[\calE,\calT(x_0)]$ is of finite accuracy, which serves to create a blurring or smearing effect on the spectrum. 
If for instance each estimate were a Gaussian random variable, centered on $d[\calE,\calT(x_0)]$ and of variance $\sigma^2$, the reconstructed spectrum would be a Gaussian blur or Weierstrass transform $W_{\sigma^2/2}$ of the underlying spectrum \eqref{ideal_profile}, and in the absence of noise this would cause spectral lines to appear as such Gaussians.
By analogy with astrophysical spectral analysis (see for instance Ref.~\cite{Fermi_LAT}), $2\sigma$ would then represent the \emph{spectral resolution} of the method in the sense that the underlying spectral structure in Eq.~\eqref{ideal_profile} should only be expected to be resolved above this lengthscale.
Spectral resolution may be improved by increasing sample size $n$ (see below).


\section{The Kolmogorov-Smirnov (KS) distance and spectrum reconstruction}
Our choice of statistical distance is the KS distance, defined as the maximum vertical distance, $d_\text{KS}[\calF,\calG]:=\sup_y|\calF(y)-\calG(y)|$, between two (cumulative or empirical) distributions $\calF(y)$ and $\calG(y)$, and so takes values in the unit interval.
We give our full justification for this choice in the SM \cite{Supp} and the relevant statistical proofs in an associated paper \cite{UP23}. 
In short though, aside from allowing for efficient calculation of unambiguous sample based estimates, properties specific to the KS distance allow us to place an analytic bound on the accuracy of such estimates, and thus also spectral resolution.
Given a pair of samples $E_1,...,E_n\sim\calE(y)$ and $T_1,...,T_m\sim\calT(y,x_0)$, the distance $d_\text{KS}[\calE,\calT(x_0)]$ is estimated by 
$d_\text{KS}[\calE_n,\calT_m(x_0)]$, where 
$\calE_n(y):=\frac{1}{n}\sum_{i=1}^{n}\mathbf{1}_{y\geq E_i}(y)$ and
$\calT_m(y,x_0):=\frac{1}{m}\sum_{i=1}^{m}\mathbf{1}_{y\geq T_i}(y)$
are empirical distribution functions that approximate $\calE(y)$ and $\calT(y,x_0)$.

\begin{figure}
     \centering
     \begin{subfigure}[b]{\linewidth}
         \centering
\includegraphics[width=\linewidth]{fig3a.pdf}
         \caption{Logarithmic scale ergodicity spectrum at 81\% fill density.}\label{log_spectrum}
     \end{subfigure}
     \hfill
     \begin{subfigure}[b]{\linewidth}
         \centering
\includegraphics[width=\linewidth]{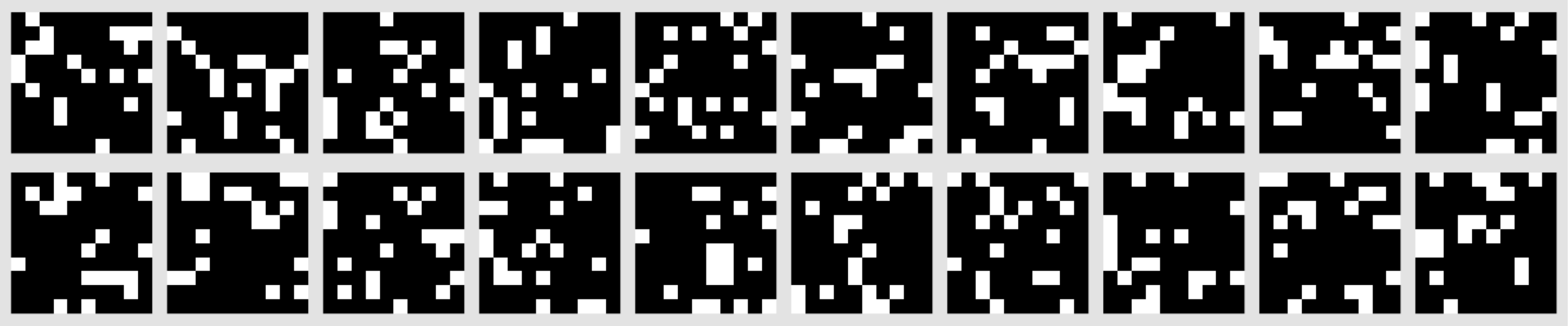}
         \caption{Random initial configs producing $d_\text{KS}[\calE_n,\calT_n(x_0)]\in [0,0.1]$.}
         \label{first_line}
     \end{subfigure}
     \hfill
     \begin{subfigure}[b]{\linewidth}
         \centering
\includegraphics[width=\linewidth]{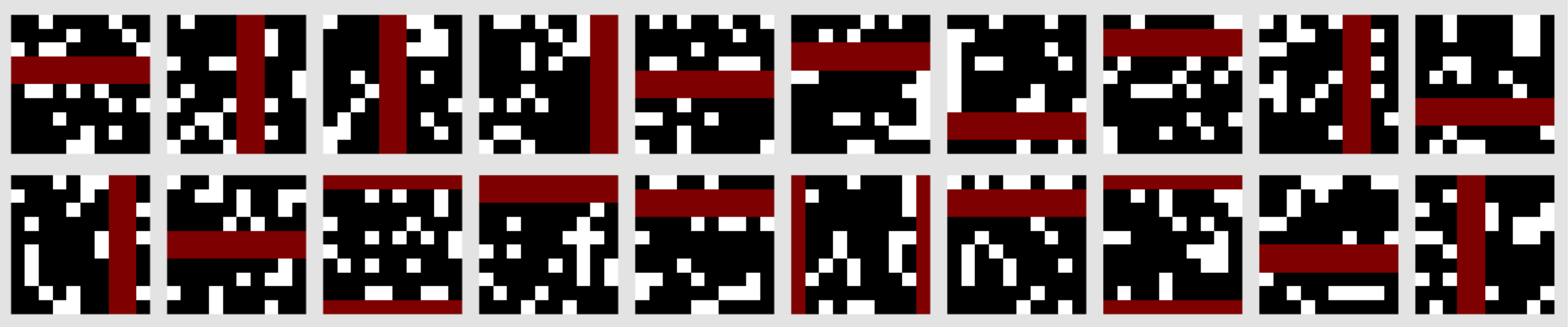}
         \caption{Random initial configs producing $d_\text{KS}[\calE_n,\calT_n(x_0)]\in[0.125,0.225]$.}
         \label{second_line}
     \end{subfigure}
     \hfill
     \begin{subfigure}[b]{\linewidth}
         \centering
\includegraphics[width=\linewidth]{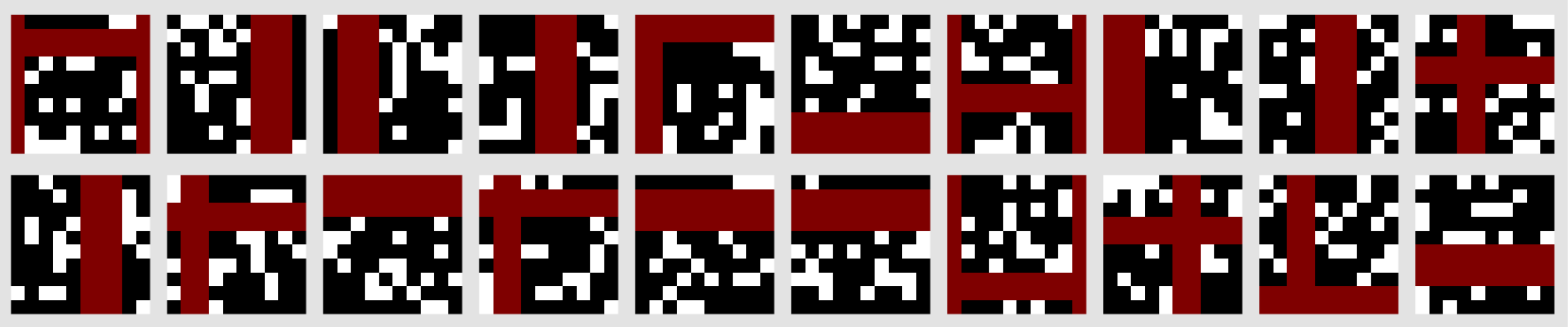}
         \caption{Random initial configs producing $d_\text{KS}[\calE_n,\calT_n(x_0)]\in[0.333,0.45]$.}
         \label{third_line}
     \end{subfigure}
\hfill
     \begin{subfigure}[b]{\linewidth}
          \centering
\includegraphics[width=\linewidth]{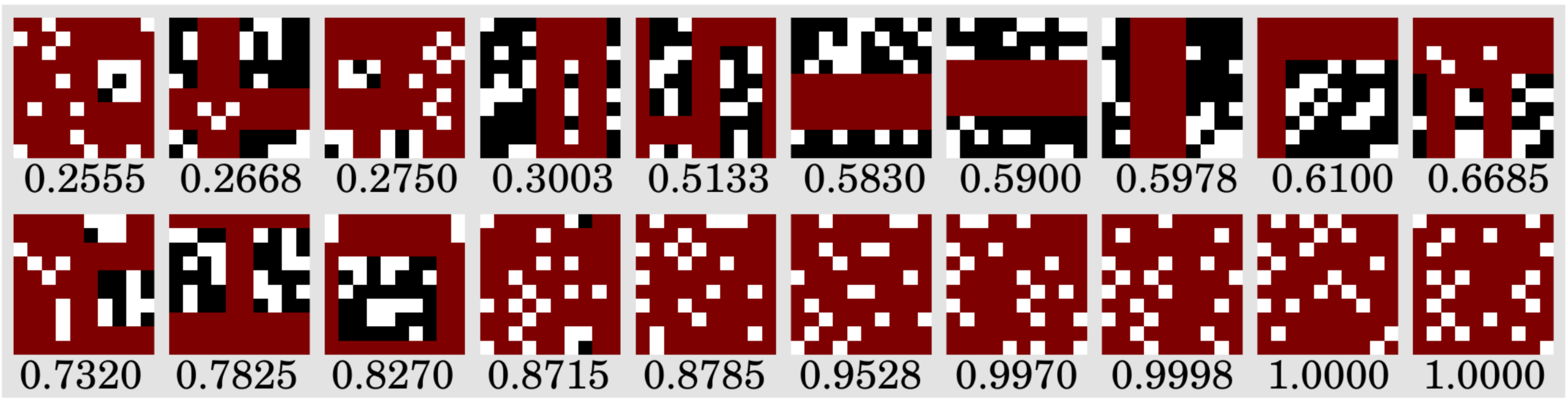}
         \caption{Rare initial configurations producing distances outside the above lines, with distances labeled below each.}
         \label{freaks}
\vspace{-8pt}
     \end{subfigure}
        \caption{Spectral lines identify distinct ergodic populations. Black: unfrozen particles. White: holes. Red: frozen particles.}
        \label{configs_figure}
\vspace{-20pt}
\end{figure}

To (re)construct a spectrum from data: 
1) Collect $n_\text{sup}$ trajectory samples of size $m$ with initial configuration $x_0$ distributed as $\mu(x_0)$.
2) Collect either a single or $n_\text{sup}$ ensemble sample(s) of size $n$. (Figs.~\ref{spectra},\ref{log_spectrum} use the latter option. See the SM \cite{Supp} for a discussion of the former.)
2) From these samples calculate $n_\text{sup}$ values of $d_\text{KS}[\calE_n,\calT_m(x_0)]$, by pairwise comparison if appropriate.
3) Create a histogram from the resulting $n_\text{sup}$ distances. 

For equal sample sizes $n=m$, as below, resolution is bounded by a $p$\% containment window of width not greater than $\beta(p) n^{-1/2}$, with $\beta(0.70)=3.63$ and $\beta(0.95)=4.55$.
Calculated values of $d_\text{KS}[\calE_n,\calT_n(x_0)]$ have at least a probability $p$ to fall in window of this width centered on $d[\calE,\calT(x_0)]$, and spectral structure should only be expected to be resolved above this scale.
Ergodic (MI or EoD) systems appear, for large $n$, as Kolmogorov's PDF $L(\sqrt{n/2}d_\text{KS}[\calE_n,\calT_n(x_0)])'$ \cite{K33_with_note_on_typo,UP23} (see Figs.~\ref{spectra},\ref{log_spectrum}).%

\section{Demonstration: Kob-Andersen (KA) model}
The KA model \cite{KA93} is a member of a class of kinetically constrained lattice glass models \cite{glassy_review1,glassy_review2_published}, the (non-)ergodicity of which has been studied in depth by Toninelli \cite{TBF05}.
In brief, a lattice is partially occupied by a fixed number of particles, with only one particle permitted to occupy each lattice site.
The dynamics comprise each particle attempting to move with rate unity to a randomly selected neighboring site, with kinetic constrainment parameterized by integer $m_\text{KA}$; each attempted move is only accepted if the particle has at most $m_\text{KA}$ occupied neighboring sites both before and after the move. 

The 2d $m_\text{KA}=2$, $L\times L$ model used in Figs.~\ref{spectra} and \ref{configs_figure} is ergodic (MI) wrt the Bernoulli product measure in the thermodynamic limit $L\to\infty$. On finite lattices MI is formally broken for particle densities $\rho\geq 2/L$ due to the potential for \emph{frozen structures} in the configuration, the simplest of which is a adjacent pair of completely filled rows or columns.
At lower densities, ergodic regions with such frozen structures are rare, and we may expect EoD and EoA to hold with high precision. 
Toninelli \cite{TBF05} describes a density dependent crossover lengthscale $\Xi[\rho]$ separating regimes where a single/many ergodic regions are sampled. 
For $L=10$ and $f$ given by the mean particle nearest neighbors, Fig.~\ref{spectra} displays spectra for densities surrounding the corresponding ``crossover density'' $\rho_c=\Xi^{-1}[L]=0.76$.
The breakdown of EoD with increasing density is apparent as a shift of the bulk of the spectrum to the right is accompanied by the appearance of additional line-like features, indicating additional ergodic regions are sampled.
Fig.~\ref{configs_figure} demonstrates how these line-like features correspond to classes of ergodic regions. 
Figs.~\ref{first_line}-\ref{third_line} show randomly selected initial configurations of trajectories producing distances in each of the first three line-like features in spectrum \ref{log_spectrum}.
The first and largest feature corresponds to no frozen structures;
the second, a single adjacent pair of frozen rows/columns (a single class of ergodic regions related by rotation/translation);
the third, two other types of frozen structure. 
We suspect due to the line shape, that this third feature may be two lines that are not resolved with the sample size $n$ used.  
Rare initial configurations that result in distances outside these lines feature more exotic frozen structures.
Fig.~\ref{freaks} shows configurations found with a dedicated search for such cases. 


\section{Concluding remarks}
After setting forth EoD as a measure of ergodicity, we have proposed a general spectroscopic method to assess and describe non-ergodicity in terms of any function of state.
The spectral profile of a system has a clear and direct interpretation through Eq.~\eqref{decomposed_spectrum}, identifying and quantifying ergodic sub-regions and/or populations. 
Application of this novel method to climate and socio-economic data may help unravel invariant sub-structures in these systems.


\bibliographystyle{modified-hunsrt}
\bibliography{/home/nick/Research/ergodicity_citations_cleaned}
\end{document}